\documentclass[aps,prl,reprint,superscriptaddress]{revtex4-2}
\usepackage{graphicx}
\usepackage{dcolumn}
\usepackage{bm}
\usepackage{color}
\usepackage{epstopdf}
\usepackage{lipsum}
\usepackage{gensymb}
\usepackage{ulem}
\usepackage{float}
\usepackage{xr}
\usepackage{sidecap}
\usepackage{mathrsfs}
\usepackage{amsmath}
\usepackage{amsthm}
\usepackage{enumerate}
\usepackage{soul}
\usepackage[symbol]{footmisc}

\begin{document}


\title{A Universal Scaling Law for Intrinsic Fracture Energy of Networks}

\author{Chase Hartquist$^{1,\dagger}$, Shu Wang$^{1,\dagger}$, Qiaodong Cui$^{2}$, Wojciech Matusik$^{2,3}$, Bolei Deng$^{4,\ast}$, Xuanhe Zhao$^{1,\ast}$\\
\normalsize{$^{1}$Department of Mechanical Engineering}\\
\normalsize{Massachusetts Institute of Technology, Cambridge, Massachusetts 02139, USA}\\
\normalsize{$^{2}$Inkbit, Medford, Massachusetts 02155, USA}\\
\normalsize{$^{3}$Computer Science and Artificial Intelligence Laboratory}\\
\normalsize{Massachusetts Institute of Technology, Cambridge, Massachusetts 02139, USA}\\
\normalsize{$^{4}$Daniel Guggenheim School of Aerospace Engineering}\\
\normalsize{Georgia Institute of Technology, Atlanta, Georgia 30332, USA}\\
\normalsize{$^\dagger$Equal contribution;}
\normalsize{$^\ast$E-mail: bolei.deng@gatech.edu, zhaox@mit.edu}
}

\date{\today}

\begin{abstract}
Networks of interconnected materials permeate throughout nature, biology, and technology due to exceptional mechanical performance. Despite the importance of failure resistance in network design and utility, no existing physical model effectively links strand mechanics and connectivity to predict bulk fracture. Here, we reveal a universal scaling law that bridges these levels to predict the intrinsic fracture energy of diverse networks. Simulations and experiments demonstrate its remarkable applicability to a breadth of strand constitutive behaviors, topologies, dimensionalities, and length scales. We show that local strand rupture and nonlocal energy release contribute synergistically to the measured intrinsic fracture energy in networks. These effects coordinate such that the intrinsic fracture energy scales independent of the energy to rupture a strand; it instead depends on the strand rupture force, breaking length, and connectivity. Our scaling law establishes a physical basis for understanding network fracture and a framework for fabricating tough materials from networks across multiple length scales.  
\end{abstract}

\maketitle

Networks ubiquitously underpin the composition of materials throughout nature and daily life, spanning from nanoscale polymers and biological materials \cite{mulla2022weak, burla2020connectivity, kassianidou2017geometry, mizuno2007nonequilibrium, chaudhuri2007reversible, vignaud2021stress} through microscale architected materials \cite{li2022mechanical, frenzel2017three, shyu2015kirigami, ling2022bioinspired, liu2021design}, synthetic tissues \cite{haines2014artificial, engelmayr2008accordion}, and structural networks \cite{bolander1998fracture, cusatis2011lattice, lilliu20033d} to macroscale fabrics and meshes \cite{wang2021structured}. The core of designing and selecting network materials that endure routine stresses in nature, technology, and daily life lies in circumventing mechanical fracture \cite{creton2016fracture,zhao2021soft,shaikeea2022toughness, romijn2007fracture}. Intrinsic fracture energy ($\Gamma_0$) -- the lowest energy required to propagate a crack per unit of created surface area -- is the key property that characterizes a material's fatigue resistance \cite{lake1967strength}. Despite its importance, no quantitative model accurately predicts the intrinsic fracture energy of networks across multiple length scales from the mechanical behavior and connectivity of its constituents. 

Griffith proposed for brittle materials that the intrinsic fracture energy is the energy to break the atomic bonds ($U_\text{bond}$) to propagate a new crack surface, i.e., $\Gamma_0/M = U_\text{bond}$, where $M$ is the areal density of bonds \cite{griffith1921vi}. Lake and Thomas extended this approach to polymers and suggested that the intrinsic fracture energy of a polymer network is the work to rupture a single layer of constituent strands ($U_\text{strand}$), i.e., $\Gamma_0/M = U_\text{strand}$ \cite{lake1967strength}. Recent experiments reveal that the Lake-Thomas model underestimates the measured intrinsic fracture energy of polymer networks by $\sim$1-2 orders of magnitude \cite{lin2021fracture, barney2022fracture, akagi2013transition, wang2021mechanism} due to nonlocal energy release \cite{deng2023nonlocal} since strands far from the crack tip relax and release energy when a crack propagates. Various models \cite{lin2020fracture, wang2023contribution} have been proposed to reconcile this discrepancy based on the assumption that $\Gamma_0/M \propto U_\text{strand}$, but none provides a physical depiction describing the fracture of networks with varying strand mechanics, network topologies, and length scales.

Here, we report a new scaling law for the intrinsic fracture energy of networks through combined simulation and experiments.

\begin{equation}
    \Gamma_0/M \propto f_f L_f
    \label{law}
\end{equation}

\noindent where $f_f$ is the strand breaking force and $L_f$ is the stretched strand length at the breaking point. We show that this result applies across multiple length scales ranging from 1 nm to 1 m for networks comprised of stretchable strands with varying single-strand force-length constitutive behaviors ranging from linear to highly nonlinear relations. We similarly demonstrate the scaling law is applicable to
a breadth of 2D and 3D network architectures, including triangular, square, hexagonal, diamond cubic, body-centered cubic, and cubic lattices. Experiments ranging from nanoscale polymer networks to macroscopic architected materials paired with simulations of networks across length scales collectively validate this scaling law. 


\begin{figure*}[t]
    \begin{center} 
    \vspace{-8pt}
    \includegraphics[width=2.05\columnwidth]{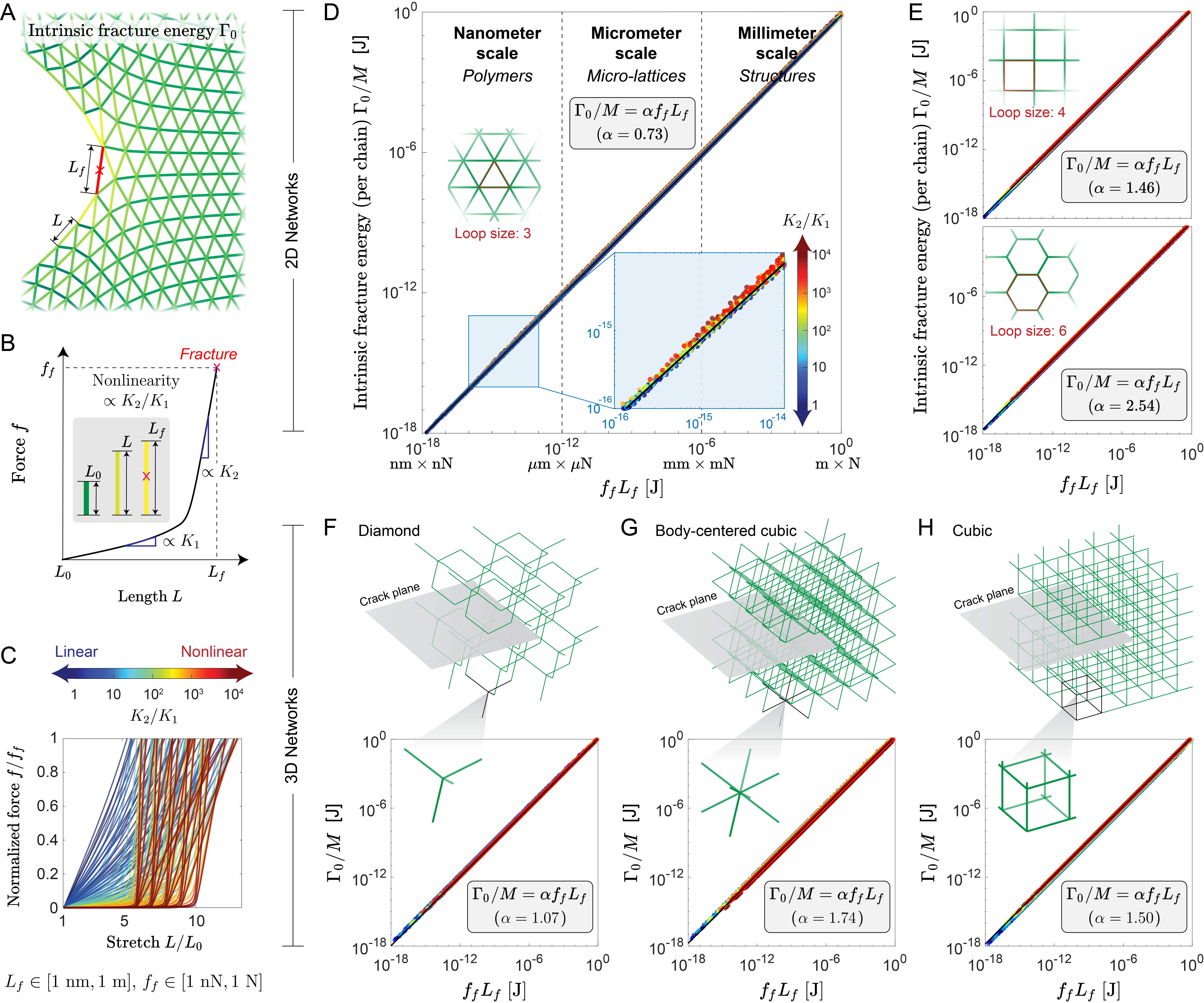}
        \caption{\textbf{Universal scaling law for intrinsic fracture energy of networks.} (\textbf{a}) Collections of identical strands are assembled into networks to measure the intrinsic fracture energy $\Gamma_0$. The crack tip of a loaded, notched specimen depicts how strand lengths ($L$, colored by stored energy) increase during loading as the bridging strand approaches its failure length ($L_f$). (\textbf{b}) The nonlinearity parameter ($K_2/K_1$) describes the strain-stiffening constitutive behavior of strands by relating the moduli of the first ($K_1$) and second regimes ($K_2$) of the force-length curve during loading to the failure force ($f_f$) and length ($L_f$). (\textbf{c}) Tuning the nonlinearity parameter ($K_2/K_1 \in$ [1, 10$^4$]), stiffness crossover length ($L_\text{x}$), failure length ($L_f \in$ [1 nm, 1 m]), and rupture force ($f_f \in$ [1 nN, 1 N]) of single strands provides a breadth of candidates for network assembly and fracture testing. (\textbf{d}) Intrinsic fracture energy (normalized by the areal strand density, $M$) scales linearly with $f_f$ and $L_f$ of single strands across all scales in a two-dimensional triangular lattice (loop size, $n_\text{loop}$ = 3), giving a prefactor $\alpha = 0.73$. (\textbf{e}) The scaling law holds for two-dimensional lattices with square ($n_\text{loop}$ = 4) and hexagonal ($n_\text{loop}$ = 6) lattices with prefactors $\alpha = 1.46$ and $2.54$, respectively. Three-dimensional networks with (\textbf{f}) diamond cubic, (\textbf{g}) body-centered cubic, and (\textbf{h}) cubic unit cells follow the scaling law with $\alpha = 1.07$, 1.74, and 1.50, respectively.}
        \label{fig1}  
    \end{center}
    \vspace{-18pt}
\end{figure*}

\section{Assembling diverse networks}

We connect the intrinsic fracture energy of networks to the constitutive behavior of individual strands by directly assembling and testing diverse networks. Mechanically identical strands with the same initial length $L_0$, terminal length $L_f$, and rupture force $f_f$ comprise each network (see Fig.~\ref{fig1}A). To describe strands with force-length behaviors varying from linear to nonlinear, we adopt the modified freely jointed chain (m-FJC) model, which relates the force $f$ and length $L$ as

\begin{equation}
    \frac{L}{L_\text{x}} = \left[\coth\left({\frac{f}{K_\text{1}}}\right)-\frac{K_\text{1}}{f}\right]\left(1+\frac{f}{K_\text{2}}\right)
    \label{m-FJC}
\end{equation}

\noindent where the two moduli $K_1$ and $K_2$ describe the stiffnesses of the force-length curve before and after the crossover length $L_{\text{x}}$.
We then iteratively assemble strands -- each with identical constitutive behavior governed by the m-FJC model -- into networks for bulk mechanical testing. Prior works introduce lattice models \cite{hrennikoff1941solution} to simulate details of brittle fracture \cite{bolander1998fracture, schlangen1992simple} and mesoscale or quasi-continuum network models to probe aspects of elastomeric fracture \cite{lei2021mesoscopic, lei2022network, ghareeb2020adaptive, yamaguchi2020topology, picu2023toughness}. Our numerical simulation performs the pure shear fracture test on the aforementioned networks by first loading a notched sample to the critical height $h_c$ where the crack propagates and then loading an unnotched sample past $h_c$ while recording the nominal stress as a function of the stretched height. The intrinsic fracture energy $\Gamma_0$ is computed as $\Gamma_0 = \int^{h_c}_{h_0}s\, dh$ and is an inherent property of sufficiently large elastic networks (see details in SI). To ensure the convergence of $\Gamma_0$, we simulate amply large networks with more than 1,000 vertical layers of strands. The goal of the numerical simulations is to connect the measured network-level intrinsic fracture energy to the preset strand-level force-length constitutive behavior. To achieve that, we systematically tune the strand failure lengths $L_f$ from 1 nm to 1 m and rupture forces $f_f$ from 1 nN to 1 N; similarly, we vary the ratio of $K_2/K_1$ as a nonlinearity parameter to match a breadth of natural and synthetic networks with behavior ranging from linear ($K_2/K_1 \approx 1$) to highly nonlinear ($K_2/K_1 \approx 10^4$) (see Fig.~\ref{fig1}C). We intentionally limit our focus to stretchable networks where the breaking stretch of each strand is greater than five (i.e., $L_f>5L_0$) to minimize geometric artifacts (see SI for a detailed explanation).

\section*{Universal scaling for network intrinsic fracture energy}

We find that all simulation results -- across strand lengths, failure forces, and nonlinearities -- follow a universal scaling law: $\Gamma_0/M\propto f_fL_f$. In Fig.~\ref{fig1}D, we plot simulated intrinsic fracture energy per strand $\Gamma_0/M$ against the product of failure force and length of the composite strands $f_f L_f$ and find that all data points collapse along a single straight line: $\Gamma_0/M = \alpha f_fL_f$, where $\alpha$ is a fitting parameter depending on the specific lattice type. While the universal scaling law holds for arbitrary strand lengths, strand breaking forces, nonlinearity parameters, and network orientations (see SI for details), the lattice topology governs the prefactor $\alpha$. The simulation yields $\alpha = 0.73$ for triangular lattices (Fig.~\ref{fig1}D), $\alpha = 1.46$ for square lattices (top of Fig.~\ref{fig1}E) and  $\alpha = 2.54$  for hexagonal lattices (bottom of Fig.~\ref{fig1}E).

Additional simulations reveal the generality of this universal scaling law to three-dimensional networks. We assemble diamond, body-centered cubic (BCC), and cubic lattices (see Fig.~\ref{fig1}F-H) from strands with the same breadth of behaviors. The universal scaling law accurately predicts intrinsic fracture energy for each three-dimensional topology. The diamond, BCC, and cubic lattices give $\alpha = 1.07$, $\alpha = 1.74$, and $\alpha = 1.50$, respectively (see Fig.~\ref{fig1}F-H). 


\begin{figure}[H]
    \begin{center} 
    \vspace{-8pt}
    \includegraphics[width=1.00
    \columnwidth]{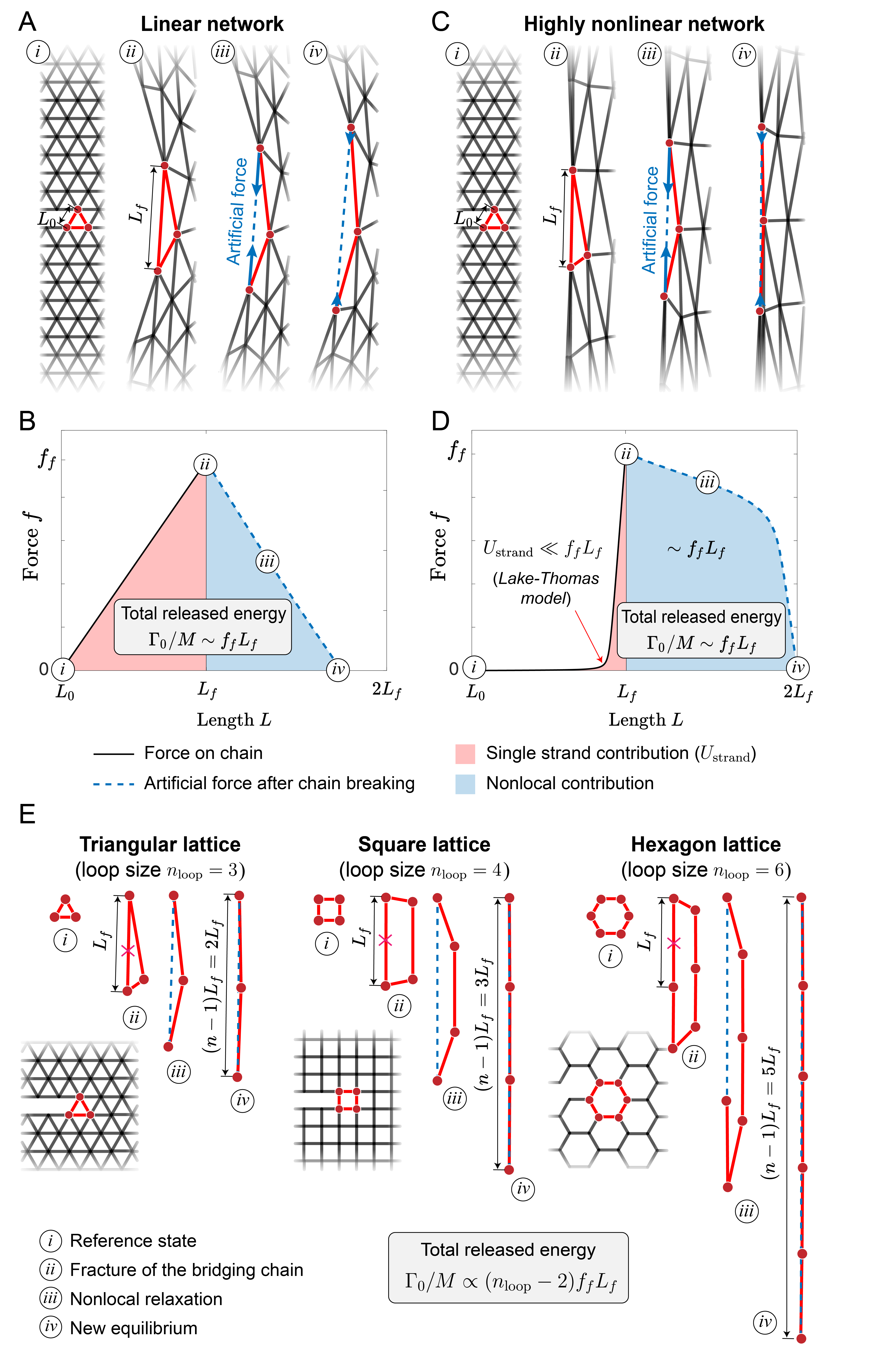}
        \caption{\textbf{Physical explanation of the scaling for intrinsic fracture energy of networks.} (\textbf{a}) A case study simulates a notched triangular network of strands with linear constitutive behavior ($K_2/K_1 \approx 1$) and loads from the undeformed state (\textit{i}) until bridging strand fracture (\textit{ii}) then quasi-statically reduces artificial forces on the ends of the broken strand (\textit{iii}) until the network reaches equilibrium (\textit{iv}). (\textbf{b}) The integration of the tracked force during strand loading (red) and nonlocal energy release (blue) as a function of length between strand ends matches the measured $\Gamma_0/M$ and scales with $f_fL_f$. (\textbf{c}) A second case study repeats the procedure for a network of strands with high nonlinearity ($K_2/K_1 \approx 10^4$) but the same $f_f$ and $L_f$. (\textbf{d}) While the single strand energy ($U_\text{strand}$, red) is much smaller than $f_fL_f$, the total integration of the single strand and nonlocal contributions counterbalance and scale with $f_fL_f$. (\textbf{e}) Simulation results depict extension from triangular ($n_\text{loop} = 3$) to square ($n_\text{loop} = 4$) and hexagonal ($n_\text{loop} = 6$) lattices for strands with high nonlinearity ($K_2/K_1 \approx 10^4$). The measured $\alpha$ parameter scales with loop size here as $\alpha \sim (n_\text{loop}-2)$. All results are numerically derived from simulations.}
        \label{fig2}  
    \end{center}
    \vspace{-18pt}
\end{figure}

\section{Physical explanation}

We next seek to understand the origins of the scaling law $\Gamma_0/M \propto f_fL_f$ by analyzing the fracture process of the bridging strand (i.e., the first breaking strand along the crack plane) and examining its energetic contributions. While loading a notched network to fracture (Fig.~\ref{fig2}A,C), the bridging strand at the crack tip stretches from $L_0$ to $L_f$ and stores elastic energy ($U_\text{strand}$) corresponding with the Lake-Thomas model (Fig.~\ref{fig2}B,D). However, instantaneously following bridging strand rupture, the network remains unbalanced and must release additional energy to reach a new equilibrium. In real networks, this additional energy can be dissipated by damping mechanisms such as viscous drag or friction. To quantitatively measure this energetic contribution, we apply a pair of artificial forces to hold the two ends of the bridging strand after it ruptures and quasi-statically reduce the magnitude of this force to zero, where the network reaches its new equilibrium. We record the value of the applied artificial forces as a function of the distance between the two ends of the ruptured strand (blue dashed line in Fig.~\ref{fig2}B,D). This integration naturally yields the energy released nonlocally by the network continuum (blue region in Fig.~\ref{fig2}B,D). Therefore, this numerical simulation quantitatively describes the energetic contributions of the single strand (red region) and network continuum (blue region) to the measured intrinsic fracture energy. 

During post-fracture relaxation, the network exerts reaction forces $\propto f_f$ across a distance $\propto L_f$ on the broken strand ends. 
For networks with linear strands (Fig.~\ref{fig2}A), the bridging strand and nonlocal contributions are on the same order of magnitude (Fig.~\ref{fig2}B, red region), so $\Gamma_0/M \propto U_\text{strand}$ still provides a reasonable approximation. However, this does not hold when applied to networks with highly nonlinear strands (Fig.~\ref{fig2}C); the single strand contribution constitutes only a small fraction of the total released energy (Fig.~\ref{fig2}D). Instead, the released energy from the network continuum dominates, which is consistent with findings from our recent work \cite{deng2023nonlocal}. Also, in networks with increasingly nonlinear strands, more strands are highly stretched, which deconcentrates stress from the crack tip. Therefore, the total measured intrinsic fracture energy of the network always scales with $f_f L_f$.

Next, we investigate how different types of networks affect the fitting parameter $\alpha$ in the universal scaling law. The key topological parameter describing a lattice during fracture is its loop size: the number of strands within the shortest closed path; our analysis includes triangular (loop size, $n_\text{loop}$ = 3), square ($n_\text{loop}$ = 4), and hexagonal ($n_\text{loop}$ = 6) lattices.
As shown in Fig.~\ref{fig2}E, the loop connected to the bridging strand in the notched sample opens when the strand breaks at $L_f$, stretching and aligning the strands within that loop. The broken strand ends in triangular, square, and hexagonal lattices migrate in this process from a distance of $L_f$ to about $2L_f$, $3L_f$, and $5L_f$, which gives relaxation lengths of $(n_\text{loop}-2) L_f$.
This result establishes a topological interpretation for the parameter $\alpha$ as $\alpha \propto (n_\text{loop}-2)$, yielding a more quantitative version of the universal scaling law:




\begin{equation}
    \Gamma_0/M \propto (n_\text{loop}-2)f_fL_f
    \label{n_minus_2}
\end{equation}

\noindent While loop size becomes more complex in 3D networks due to structural intricacies (i.e., strands with multiple adjoining loops, etc.), effective loop sizes can be found which match this result.


\begin{figure}[t]
    \begin{center} 
    \vspace{-8pt}
    \includegraphics[width=1.00
    \columnwidth]{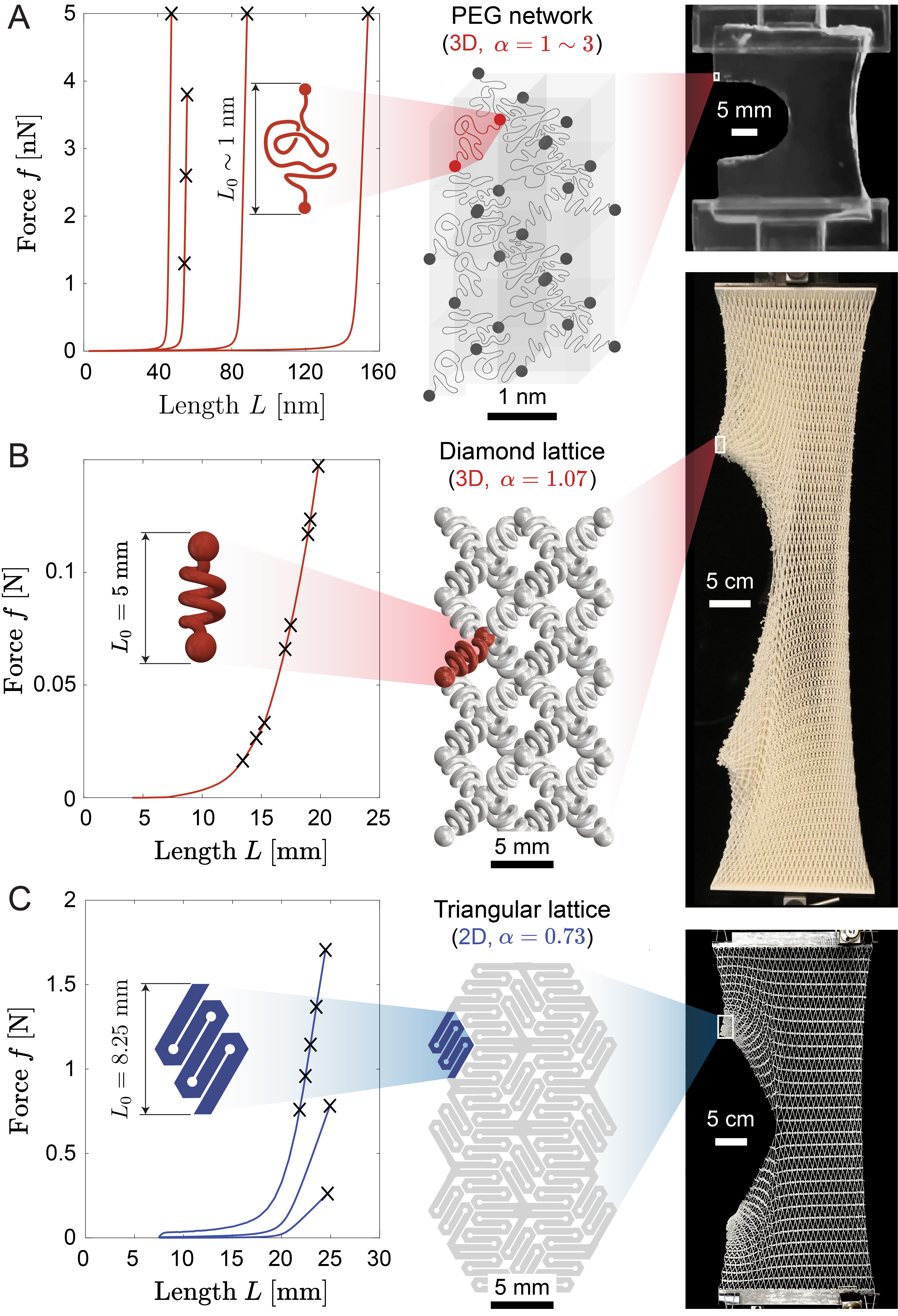}
        \caption{\textbf{Experimental networks}. (\textbf{a}) Nanoscale three-dimensional poly(ethylene glycol) end-linked polymer networks, (\textbf{b}) macroscale three-dimensional architected diamond lattices, and (\textbf{c}) macroscale two-dimensional architected triangular networks displaying various single strand force-length behaviors ($f_f$, $L_f$, and $K_2/K_1$) are assembled to measure intrinsic fracture energy via the pure shear fracture test on notched (rightmost images) and unnotched samples to validate the universal scaling law.} 
        \label{fig3}  
    \end{center}
    \vspace{-18pt}
\end{figure}


\begin{figure*}[t]
    \begin{center} 
    \vspace{-8pt}
    \includegraphics[width=1.6
    \columnwidth]{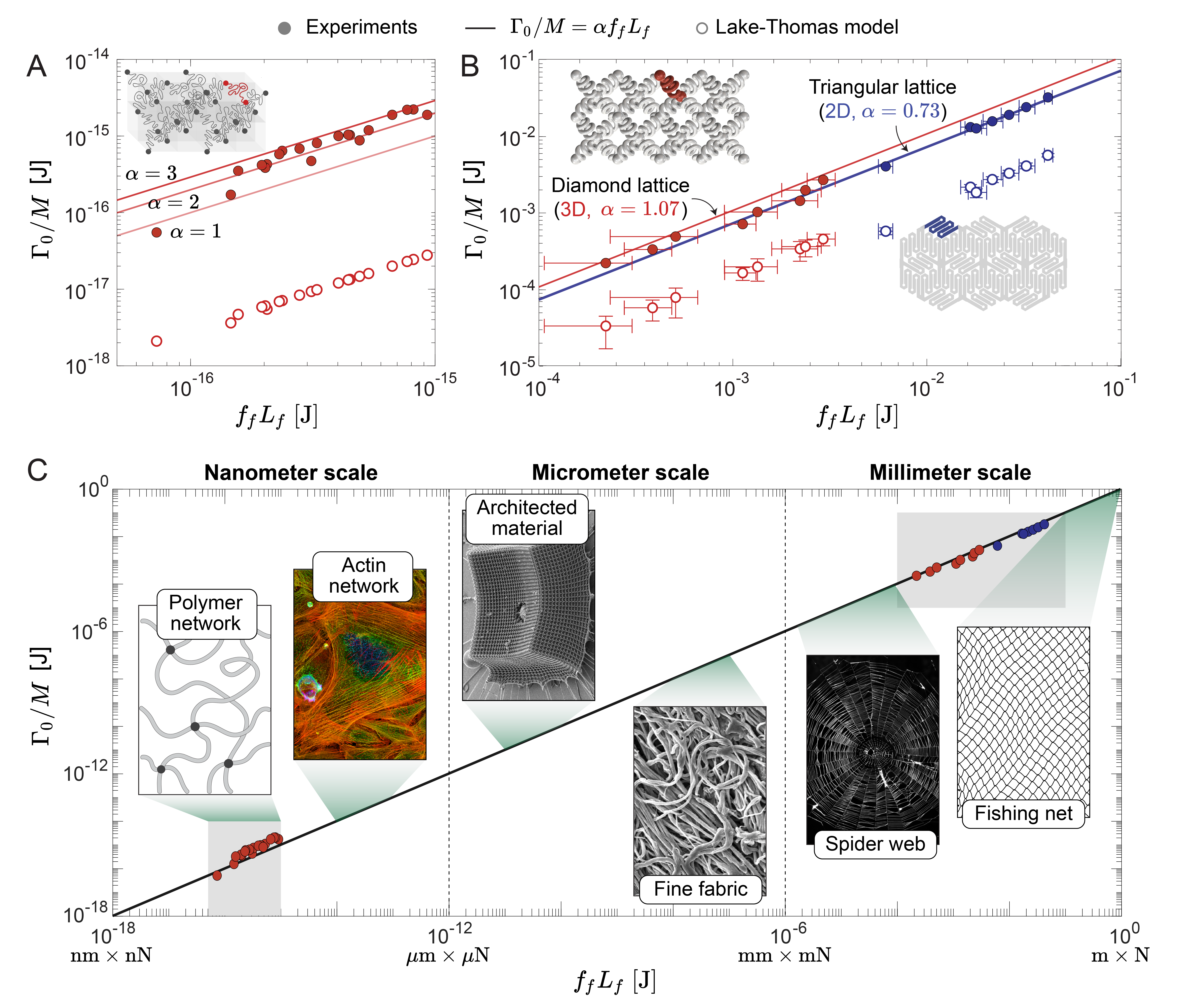}
        \caption{\textbf{Intrinsic fracture energy in experimental networks.} Experiments validate the universal scaling law in predicting the measured intrinsic fracture energies of networks ranging from (\textbf{a}) nanoscale three-dimensional PEG polymer networks from the literature to (\textbf{b}) macroscale two- and three-dimensional architected networks with various single-strand behaviors. The Lake-Thomas model prediction deviates by $\sim$1-2 orders of magnitude. (\textbf{c}) Experimental results show the applicability of the universal scaling law to a vast range of materials across scales, including polymer networks, biological networks (image courtesy of Howard Vindin), architected materials \cite{portela2021supersonic}, textiles \cite{park2018ultrafine}, spider webs (image courtesy of Chen-Pan Liao), and nets (image courtesy of Nikodem Nijaki).}
        \label{fig4}  
    \end{center}
    \vspace{-18pt}
\end{figure*}

\section{Experimental verification}
We illustrate the versatility and consistency of this universal scaling through experimental validation across a spectrum of networks, ranging from polymers composed of strands at the nanometer level to architected materials composed of strands at the millimeter level. For nanoscale polymer networks, we collect intrinsic fracture energy measurements from across the literature for tetra-poly(ethylene glycol) (PEG) hydrogels since they possess relatively homogeneous networks \cite{sakai2008design} (see Fig.~\ref{fig3}A). Degree of polymerization $N$ between crosslinks tunes the failure length of polymer strands ($L_f \sim N$) \cite{akagi2013transition} while mechanophores embedded in the backbone tune the rupture force of strands $f_f$ \cite{wang2021mechanism,lin2021fracture}. For macroscale architected materials, we fabricate two- and three-dimensional networks from folded and spring-shaped strands, respectively, with various strain-stiffening $K_2/K_1$ behavior by controlling the transition between compliant unfolding ($K_1$) and stiffer material stretching ($K_2$) during loading. Two-dimensional triangular lattices of repeating strands are laser cut (model: Epilog Laser Fusion Maker 12) from polyester (0.001” thick) and polyacetal (0.003”, 0.005” thick) films (McMaster-Carr) (Fig.~\ref{fig3}B). Three-dimensional diamond lattices of repeating strands are modeled in a commercial 3D modeling software (SolidWorks, Dassault Systems) and 3D printed (Inkbit Vista, Inkbit) using a thiol-ene polyurethane elastomer (TEPU30A, Inkbit) \cite{buchner2023vision} (Fig.~\ref{fig3}C). Intrinsic fracture energy is calculated after loading an unnotched specimen in pure shear to obtain the force-stretch behavior and extending notched specimens in pure shear to the critical height $h_c$ where bridging strands reach $L_f$.

These experimental networks cover a vast spectrum of single-strand nonlinearity parameters ($K_2/K_1$ from 20 to $1.8 \times 10^4$), breaking forces ($f_f$ from 1.3 nN to 1.7 N), and breaking lengths ($L_f$ from 47.4 nm to 25.1 mm). Across this range, the proposed universal scaling law predicts the experimentally measured polymer (Fig.~\ref{fig4}A) and architected (Fig.~\ref{fig4}B) network intrinsic fracture energy, while the Lake-Thomas model prediction underestimates all measurements by orders of magnitude. The PEG polymer network results match well with $\alpha$ between 1-3 and scale appropriately (Fig.~\ref{fig4}A). The $\alpha$ value of 0.73 from triangular lattice simulations shows strong agreement with the experiments for two-dimensional architected triangular networks; similarly, the value of 1.07 from diamond lattice simulations matches well with experiments for three-dimensional architected diamond networks (Fig.~\ref{fig4}B). 
The Lake-Thomas model consistently underestimates the experimental results, but the deviation varies in extent from one to two orders of magnitude. Overall, experimental agreement of nanoscale polymer networks and macroscopic architected materials to the scaling law synergistically promote its universal applicability in predicting diverse network fracture (Fig.~\ref{fig4}C).

\section{Discussion}
Notably, for polymer networks, the scaling of the Lake-Thomas model with respect to strand length coincides with the universal scaling law presented here. The Lake-Thomas model suggests that $\Gamma_\text{0}/M \sim U_{\text{strand}} \sim N$, where $N$ is the number of bonds within a polymer between junctions. Our result $\Gamma_0/M \propto f_f L_f$ leads to the same scaling for polymer networks since $L_f \sim N$. Therefore, both the Lake-Thomas model and the universal scaling law predict that $\Gamma_0/M \propto N$. The key distinction highlighted here is that the measured intrinsic fracture energy is on the order of $f_fL_f$ and can be much larger than $U_{\text{strand}}$. This observation reconciles results from the literature which claim that the Lake-Thomas model can effectively match measured trends while deviating in magnitude \cite{akagi2013fracture}.

The findings shared in this work parallel a separate phenomenon studied in linear elastic fracture mechanics based on Griffith theory called lattice trapping \cite{thomson1971lattice}. Lattice models implementing atomistic force laws to approximate fracture of atomically bonded materials expose a mismatch in the measured critical energy release rate for crack propagation and the surface energy dissipated by breaking bonds \cite{bernstein2003lattice}. This mismatch occurs because the lattice structure exhibits an energy barrier which traps the network in a local energy minimum even though the fractured state is the global energy minimum. The nonlocal contribution to intrinsic fracture energy parallels this lattice trapping effect. For real crystalline materials, this energy barrier typically remains small enough such that the combination of thermal energy, dislocations, grain structure, etc. cause the crack to overcome the barrier to fracture within relevant timescales \cite{curtin1990lattice, sinclair1975influence}. For networks described here, this energy barrier can become extremely steep. We contend that the crack would likely not overcome this barrier on relevant timescales due to thermal fluctuations or structural defects. The true impact of energy fluctuations on lattice trapping across all networks and length scales remains an open question. We propose that the full local and nonlocal effects must be considered together to capture the intrinsic fracture energy measured in networks studied here.

Networks manifest in nature due to exceptional resistance to failure under harsh loading conditions. Designing materials that mechanistically resist fracture requires an understanding of the hierarchical connection between strand mechanics, network connectivity, and macroscopic properties. Here we reveal a simple universal scaling that unifies the fracture of networks across many length scales, strand mechanical nonlinearities, and lattice topologies. Advancing a crack through a network requires local energy dissipation through breaking the bridging strand and nonlocal energy release through opening the adjoining loop to rebalance. The physical process counterintuitively produces a measured intrinsic fracture energy that scales with geometry through $n_\text{loop}$ and single strand mechanics through $f_fL_f$ instead of $U_{\text{strand}}$. This finding provides a foundational mechanism for interpreting and designing networks as tough materials. For instance, nanoscale polymer strands garner toughness in natural, biological, and synthetic networks by synergistically achieving high deformations and rupture forces. Similarly, animals such as bees and spiders leverage connectivity to resist honeycomb and web fracture. This universal scaling law not only elucidates the beauty of existing network structures but informs future design of lattices in metamaterials, textiles, and beyond. 



\bibliography{References}
\bibliographystyle{unsrt}


\section{Methods}


\subsection{Mathematical model for numerical simulation}
We model networks as systems of connected nonlinear springs with constitutive force $f$ and length $L$ relations characterized by the modified freely jointed chain model in Eq.~(\ref{m-FJC}). 
This formula yields the constitutive law we apply to each nonlinear spring in the simulation. To capture strand rupture, a breaking force $f_f$ and length $L_f$ are prescribed to the spring. We vary the range of $f_f$ from 1 nN to 1 N, $L_f$ from 1 nm to 1 m, and $K_2/K_1$ from 1 to 3.0 $\times$ 10$^4$ to broadly describe networks across scales. High nonlinearity parameters capture the extreme strain-stiffening behavior measured using single molecule force spectroscopy for common polymers such as poly(acrylic acid) \cite{li1999single}, poly(vinyl alcohol) \cite{li1998single}, polyisoprene \cite{zhang2002nano}, poly(acryl amide) and poly(N-isopropyl acrylamide) \cite{zhang2000single}, poly(dimethylacrylamide) and poly(diethylacrylamide) \cite{wang2002force}, and poly(ethylene glycol), whose nonlinearity parameter reaches upwards of 1.8 $\times$ 10$^4$ \cite{oesterhelt1999single}. The breaking length $L_f$ is notably selected in simulations to be at least five times the initial length (i.e., $L_f > 5 L_0$) to limit geometric artifacts.

We describe the lattice deformation of general 3D networks consisting of $n$ nodes and $e$ edges through their coordinates $(x_i,y_i,z_i)$, where $i = $ 1$,...,n$. Two matrices store the node coordinates and their respective connectivities in MATLAB. The total system energy at each loading step is expressed as the sum of the elastic energy stored in each edge or spring as,
\begin{equation}
    U_\text{total} = \sum_{i,j} \int_1^{\lambda_{ij}} f(\lambda')d\lambda',
\end{equation}
where $\lambda_{ij}$ is the stretch of the edge connecting node $i$ with $j$:
\begin{equation}
    \lambda_{ij} = r_0^{-1}\sqrt{(x_i - x_j)^2 + (y_i - y_j)^2 + (z_i - z_j)^2}.
\end{equation}

Minimizing $U_\text{total}$ numerically provides the coordinates of each node $(x_i,y_i,z_i)$ as solved by equating
 \begin{equation}
 \label{ge_static}
    \frac{\partial U_\text{total}}{\partial x_i} = 0,\;\frac{\partial U_\text{total}}{\partial y_i} = 0,\;\frac{\partial U_\text{total}}{\partial z_i} = 0,
\end{equation}
using Newton's method in MATLAB. Additionally, a broken edge between nodes $i$ and $j$ is detected when $\lambda_{ij}>\lambda_f$ and removed by deleting the corresponding entries of the connectivity matrices.

Clamped boundary conditions are prescribed in the simulation to the top and bottom surface to quasi-statically stretch the sample from the initial height $h_0$ to a final height $h$ in the $y$-direction. The displacement boundary condition is applied in the simulation on the top and bottom nodes as,
\begin{equation}
    \label{bc1}
    \begin{split}
        y_i &= h,\,\text{for }i\in \text{top nodes},\\
        y_i &= y_i^0,\,\text{for }i\in \text{bottom nodes},
    \end{split}
\end{equation}
where $y_i^0$ denotes the initial $y$ position of the $i$-th nodes. 

The sample width $w_0$ in the $x$-direction is set to twice the height $h_0$ in all simulations as $w_0 = $ 2$h_0$. We fix the $x$-displacement to enforce a pure shear loading condition and limit boundary effects via 
\begin{equation}
    \label{bc2}
    \begin{split}
        x_i &= x_i^0,\,\text{for }i\in \text{left nodes},\\
        x_i &= x_i^0,\,\text{for }i\in \text{right nodes},
    \end{split}
\end{equation}
where $x_i^0$ denotes the initial $x$ position of the $i$-th nodes. Eqs.~(\ref{ge_static}-\ref{bc2}) form a boundary value problem that can be solved numerically. 

\subsection{Quasi-static solver}

The node coordinates $(x_i,y_i,z_i)$ fully describe the state of the system, so all variables can be rewritten as vectors:
\begin{equation}
    \mathbf{X} = [x_1,y_1,z_1,x_2,y_2,z_2, ... , x_n, y_n, z_n]^T.
\end{equation}
The 3$n$ by 1 vector $\mathbf{X}$ contains all necessary information to describe the lattice deformation. The nonlinear system of equations described in Eqs.~(\ref{ge_static}-\ref{bc2}) is solved to obtain $\mathbf{X}$ and can be written generally as,
\begin{equation}
\label{ge_static_vector}
    \mathbf{F}(\mathbf{X}) = \mathbf{0}.
\end{equation}
Note that the equation above presents the same governing equations depicted in Eq.~(\ref{ge_static}). 

The solver implements Newton's method to solve the governing equation (Eq.~(\ref{ge_static_vector})). The generalized Newton's method is to find a root of a functional \textbf{F} defined in a Banach space. In this case, the formulation is
\begin{equation}
\label{newton_J}
\textbf{X}_{l+1}=\textbf{X}_{l}-\Big[\textbf{J}(\textbf{X}_{l})\Big]^{-1}\textbf{F}(\textbf{X}_{l}),\\
\end{equation} 
\noindent where $\textbf{J}(\textbf{X}_{l})$ is the Jacobian matrix of the function $\textbf{F}$ at $\textbf{X}_{l}$, and $l$ is the iteration number. Instead of computing the inverse of this matrix, one can save time by solving the following system of linear equations:
\begin{equation}
\textbf{J}(\textbf{X}_{l})\left(\textbf{X}_{l+1}-\textbf{X}_{l}\right)=-\textbf{F}(\textbf{X}_{l}).\\
\end{equation}
\noindent Starting with an initial guess $\textbf{X}_0$, the next approximate solution $\textbf{X}_l$ is obtained iteratively. The method ends when $\|\textbf{X}_{l+1}-\textbf{X}_{l}\|<\delta$, where $\delta$ is a defined accuracy requirement. 

The quasi-static simulation divides the loading process into $P$ steps to gradually stretch the network. It obtains the system state $\textbf{X}^{(p)}$ by solving Eq.~(\ref{ge_static_vector}), where $p =$ 1$,..,P$, at each step. To accelerate convergence of Newton's method, the solution of the current step provides the initial guess for the upcoming step.

\subsection{Post-rupture artificial force decay simulation}
We adapt the simulation protocol to explore the physical explanation for the universal scaling law by tracking and relaxing the nodes connecting the first bridging strand following rupture. A notched network is first loaded to the critical height $h_c$ where the first bridging strand breaks. The bulk boundary conditions are fixed for the remainder of the simulation. Instead of breaking the bridging strand and equilibrating the system, the strand is replaced by an approximately infinitely stiff spring ($K_\text{spring} \gg K_2 \ge K_1$). At the initial step, the simulation fixes the length of the stiff spring to match the length of the broken strand ($L_\text{spring} = L_f$), obtains the next system state $\textbf{X}^{(P+1)}$ by solving Eq.~(\ref{ge_static_vector}), and stores the pair of opposing artificial forces required to preserve equilibrium on each node. For the remaining steps, the algorithm incrementally lengthens the spring, equilibrates the system state, and stores the new spring force readout. The iterative procedure concludes once the measured spring force reaches a small tolerance of zero or the spring reaches a predetermined terminal length.

\subsection{Coarse-graining simulation procedure}

We simulate networks on the order of thousands of layers to ensure convergence of samples with high nonlinearity parameters (see SI for details on convergence). Networks are coarse-grained far from the crack tip for computational efficiency (see SI for a case study on the computational limits). The coarse-grained method reconstructs large networks with drastically fewer degrees of freedom (DOFs). Near fracture of the bridging strand, the network is most inhomogeneous near the crack tip but becomes more homogeneous with increasing distance. Since strands far from the crack tip do not substantially vary in their local neighborhoods, a coarse lattice can equivalently describe their continuum-level mechanical response. For instance, a 2D triangular network with $h_0 =$ 100 layers and $w_0 =$ 200 layers can be reconstructed with incremental levels of coarse lattices moving radially outwards from the crack tip (see 
the SI figure on coarse-graining). The relative stiffness of coarse-grained strands is prescribed proportionally to length (represented by line thicknesses in 
the SI figure on coarse-graining) to ensure the coarse-grained neighborhoods maintain the same bulk mechanical performance as the full network. In notched samples, levels incrementally coarsen with increasing distance in the $x$- and $y$-directions from the undeformed crack tip. A full two-dimensional network possessing 23,057 nodes can therefore be coarse-grained with this scheme to contain only 1,047 nodes (see 
SI figure on coarse-graining). Direct comparisons indicate that the coarse-grained model accurately predicts the critical stretch at which the first strand breaks (see SI for a detailed comparison). While the coarse-grained model cannot accurately capture the full fracture process, it yields an accurate measure of $h_c$ in the pure shear fracture test.

The coarse-grained triangular networks used for all simulations contain a size of $h_0 =$ 4,000 layers by $w_0 =$ 8,000 layers, with a total of 44,847 nodes and 89,694 DOFs (see 
SI figure on coarse-grained sample size). Note that an equivalent full network requires about 40 million nodes; the coarse-graining scheme decreases the required number by 99.9$\%$. Each iteration of Newton's method -- which includes assembling the Jacobian matrix and solving Eq.~(\ref{newton_J}) -- typically costs a few seconds. The full fracture simulation of a two-dimensional $h_0 =$ 4,000 layer network typically takes under 20 minutes to complete on a standard desktop with Intel Core i9- 12900K.


\subsection{Fabrication of architected networks}
Two-dimensional networks are fabricated by laser cutting polyester (12" $\times$ 12" $\times$ 0.001") and polyacetal (12" $\times$ 12" $\times$ 0.003", 0.005") sheets (McMaster-Carr part numbers: 7594T11 and 5742T11) with a laser cutter (Fusion Maker 12, Epilog Laser). The triangular network strand pattern is designed (CorelDRAW, Corel Corporation) with 28 vertical layers of repeating units with 60 strands per layer. Each strand has a ``zigzag'' structure that unfolds to provide an initial compliant bending regime then deforms the material to provide a final stiff stretching regime \cite{jang2015soft}. This large discrepancy between stiffness enables high values and tunability of the nonlinearity parameter $K_2/K_1$ from the m-FJC model. The distance between the laser head and the acetal film is calibrated before cutting to ensure sharp focus. Cutting parameters are selected to be 10\% for laser power, 10\% for frequency, and 100\% for speed. Four identical samples in total are cut to perform each measurement of energy release rate. For each sample, four 1/16" acrylic sheets are cut and glued on either side of an uncut portion at the top and bottom of the sample to act as a rigid boundary, which is clamped onto the mechanical testing machine.

For three-dimensional networks, single spring-shaped strands are designed and parameterized using commercial 3D modeling software (Solidworks, Dassault Systems). Strands are assembled spatially into a diamond cubic unit cell and joined at strand ends via spherical nodes. Unit cells comprised of 16 strands are patterned into a 16$\times$16$\times$8 array. The resulting lattice contains 2,048 unit cells and 32,768 single strands. Rectangular plates are joined to the top and bottom faces of the network. The assembled components are joined and resized such that single strands are 5 mm in length, giving a bulk height of 186 mm for the 16 unit cells. The resulting network is saved as a sterolithography file, exported, and 3D printed on a vision controlled jetting system (Inkbit Vista, Inkbit) using a thiol-ene polyurethane elastomer (TEPU30A, Inkbit). For each sample, four acrylic sheets and cut and glued to the top and bottom plates to form a rigid boundary with a vertical protrusion to be clamped onto the testing machine.

\subsection{Experimental measurement of architected networks}
Pure shear fracture tests are performed on two-dimensional networks using a ZwickiLine materials testing machine (2.5 kN load cell, Zwick/Roell). To measure fracture energy $\Gamma_0$, a uniaxial extension test is first performed on a pristine sample at a constant loading rate of 100 mm/min. Using the experimentally obtained uniaxial response, we inversely identify the effective force-length curve for each strand such that the simulation results match experiments. To validate this approach, five individual 0.005" polyacetal strands are laser cut and tested uniaxially. The average force-length curve measured experimentally matches the profile obtained by the inverse method. The inverse method is then used for 0.003" polyacetal and 0.001" polyester two-dimensional networks. For the remaining three samples, we introduce an identical crack with length $\sim w_0/$2. Uniaxial tensile tests are performed on the three notched samples at a loading rate of 100 mm/min. Since the rupture of strands is uncontrolled when a notched sample is loaded to the fracture event, we preset the critical stretched height $h_c$ and consider the bridging strands to ``rupture'' when the whole network reaches that applied height. The energy release rate of the pristine sample at $h_c$ is measured and recorded to mark the intrinsic fracture energy $\Gamma_0$ of the network. The rupture length $L_f$ of the bridging strand is measured using calipers when the notched sample reaches $h_c$, and the rupture force $f_f$ is interpolated from the single strand force-length curve.

Three-dimensional pure shear tests are performed using a single-axis Instron universal testing machine (500N load cell, Instron 5566) at a constant loading rate of 1 mm/s. Prior to mechanical loading, rigid acrylic mounts are glued to the rectangular plates printed on the top and bottom faces of the sample. Mounts are then fixed via mechanical grippers within the testing apparatus. The fracture energy $\Gamma_0$ is calculated by the same procedure outlined for the 2D networks. Note that crack of width $w_0/$2 is cut through all layers in the thickness direction. The rupture length $L_f$ of the bridging strand is measured based on snapshots of experimental recordings. At each loading state, we measure five bridging strands in the thickness direction and take their average as the current $L_f$. Note that due to gravity, the strands fall on each other at the original height of 186 mm. The network is not fully opened until it is been stretched to 350 mm. To eliminate the effect of gravity, we set the measured force to 0 N until the network is stretched to 350 mm.

\end{document}